\documentclass{llncs}
\usepackage{time}
\usepackage{cases}

\newcommand{\p}{{\cal P}}

\usepackage{ifpdf}
\ifpdf                             
\usepackage[pdftex]{graphicx}      
\usepackage[pdftex]{hyperref}
\else 
\usepackage{epsfig}
\usepackage[hypertex]{hyperref}
\fi 
\hypersetup{
    pdfauthor   = {Hristo Djidjev},
    pdftitle    = {A fast multilevel algorithm for graph clustering and
                    community detection},
    colorlinks  = true,
    pdfkeywords = {graph clustering, graph partitioning, community detection,
                    multilevel optimization}, }

\usepackage{fancyhdr}

\begin{document}

\title{\bf A scalable multilevel algorithm for graph clustering and community
structure detection}
\author{Hristo N.~Djidjev \inst{1} \thanks{This work has been
supported by the Department of Energy under contract
W-705-ENG-36.}
\institute{Los
Alamos National Laboratory, Los Alamos, NM 87545
}}

\maketitle
\begin{abstract}
One of the most useful measures of  cluster quality is the
modularity of the partition, which measures the difference
between the number of the edges joining vertices from the
same cluster and the expected number of such edges in a
random (unstructured) graph. In this paper we show that
the problem of finding a partition maximizing the
modularity of a given graph $G$ can be reduced to a
minimum weighted cut problem on a complete graph with the
same vertices as $G$. We then show that the resulted
minimum cut problem can be efficiently solved with
existing software for graph partitioning and that our
algorithm finds clusterings of a better quality and much
faster than the existing clustering algorithms.
\end{abstract}

\section{Introduction}
One way to analyze and understand the information
contained in the huge amount of data available on the WWW
and the relationships between the individual items is to
organize them into "communities," maximal groups of
related items. Determining the communities is of great
theoretical and practical importance since they correspond
to entities such as collaboration networks, online social
networks, scientific publications or news stories on a
given topic, related commercial items, etc. Communities
also arise in other types of networks such as computer and
communication networks (the Internet, ad-hoc networks) and
biological networks (protein interaction networks, genetic
networks).

The problem of identifying communities in a network is
usually modeled as a \emph{graph clustering} (GC) problem,
where vertices correspond to individual items and edges
describe relationships. Then the communities correspond to
clusters with a lot of edges between vertices belonging to
the same subgraph (called \emph{in-cluster} edges) and
fewer edges between vertices from different subgraphs
(called \emph{between-cluster} edges). The GC problem has
been intensively studied in the recent years in relation
to its applications in the analysis of networks. Girvan
and Newman propose in \cite{GN2002}, \cite{NewmanGirvan}
algorithms based on the \emph{betweenness} of the edges of
a graph, a characteristic that measures the number of the
shortest paths in a graph that use any given edge. In
\cite{Newman2004} Newman describes an algorithm based on a
characteristic of clustering quality called
\emph{modularity}, a measure that takes into account the
number of in-cluster edges and the expected number of such
edges. (We formally define and discuss modularity in more
detail in the next section.) A faster version of the
algorithm from \cite{Newman2004} was described by Clauset
\emph{et al.} in \cite{CNM}. Several algorithms have been
proposed based on the eigenvectors of the graph Laplacian,
e.g., \cite{WhiteSmyth}, \cite{NewmanEigenvector}. In all
previous cases the algorithms reported in the literature
are either not fast enough, or are inaccurate.

In this paper we will describe a new approach for GC that
uses our newly discovered relationship between the GC and
the minimum weighted cut problems. The \emph{minimum
weighted cut} (MWC) problem is, given a graph $G=(V,E)$
with real weights on its edges, find a partition of $V$
such that the set of all edges of $G$ that join vertices
from different sets of the partition, called a \emph{cut}
of the partition, is of minimum weight. GC looks
related to the MWC problems since, in a good quality
clustering, the weight of the edges between different sets
of the partition (the cut) should be small compared to the
weight of the edges inside the sets. But the MWC problem
can not be directly applied to solve the GC problem since
it does not take into account the sizes of the subgraphs
induced by the cut (e.g., it is likely that the minimum
cut will consist of the edges incident to a single vertex). There
are some minimum cut based clustering algorithms, e.g.,
\cite{FlakeTarjan}, that use maximum flow computations
combined with heuristics, but they are typically slower
than modularity based algorithms, e.g. \cite{CNM}, and,
moreover, they cannot determine the optimal number of
clusters and, instead, construct a hierarchical
decomposition of the set of all vertices of the graph.

In this paper we prove that the problem of finding a partition of a
graph $G$ that maximizes the modularity can be reduced to
the problem of finding a MWC of a weighted complete graph
on the same set of vertices as $G$. We then show that the
resulting minimum cut problem can be solved by modifying
existing fast algorithms for graph partitioning. We
demonstrate by experiments that our algorithm has
generally a better quality and is much faster than the
best existing GC algorithms.

\section{Our clustering algorithm}

\subsection{Graph clustering as a minimum cut problem}
\label{sec:mincut}

As there is no formal definition of clustering and what
the clusters of a given graph are, in general it is not
possible to determine if a certain partition is the
"correct" clustering or which of two alternative
partitions of a graph corresponds to a better clustering.
For that reason, researchers have used their intuition to
define measures for cluster quality that can be used for
comparing different partitions of the same graph. One such
measure, introduced in \cite{NewmanGirvan,Newman2003},
which has received considerable attention recently, is the
\emph{modularity} of a graph. Given an $n$-vertex $m$-edge
graph $G=(V(G),E(G))$ and a partition $\cal P$ of $V(G)$
into $k$ subsets (clusters) $V_1,\dots,V_k$, the
modularity $Q({\cal P})$
of $\cal P$ is a number defined as %
$$Q({\cal P}) = \frac{1}{m}\sum_{i=1}^k(|E(V_i)|-{\rm Ex}(V_i,{\cal G})),$$
where $E(V_i)$ is the set of all edges of $G$ with
endpoints in $V_i$ and ${\rm Ex}(V_i,{\cal G})$ is the
expected number of such edges in a random graph with a
vertex set $V_i$ from a given random graph distribution
$\cal G$. $Q({\cal P})$ measures the difference between
the number of in-cluster edges and the expected value of
that number in a random (e.g., without cluster structure)
graph on the same vertex set. Larger values of $Q({\cal
P})$ correspond to better clusterings.

Having the definition of $Q({\cal P})$, we can formulate
the clustering problem as finding a partition ${\cal
P}=\{V_1\cup\dots\cup V_k\}$ of $V(G)$ such that
\begin{equation}
\sum_{i=1}^k(\;|E(V_i)|-{\rm Ex}(V_i,{\cal G}))\rightarrow
\max. \label{eq:Q1}
\end{equation}

Clearly $$\max_{\cal P}\{\;\sum_{i=1}^k(\;|E(V_i)|-{\rm
Ex}(V_i,{\cal G})\;)\}$$ $$= -\min_{\cal
P}\{\;-\sum_{i=1}^k(\;|E(V_i)|-{\rm Ex}(V_i,{\cal G})\;)\}$$
$$= -\min_{\cal P}\{\;(|E(G)|-\sum_{i=1}^k\;|E(V_i)|\;)
-(|E(G)|-\sum_{i=1}^k\;{\rm Ex}(V_i,{\cal G})\;)\} $$ $$=
-\min_{\cal P}\{\;|{\rm Cut}({\cal P})|-{\rm ExCut}({\cal
P},{\cal G})\},$$ where ${\rm Cut}({\cal P})$ is defined as
the cut of ${\cal P}$ and ${\rm ExCut}({\cal P},{\cal G})$
the expected value of ${\rm Cut}({\cal P})$ for a random
graph from $\cal G$.

Hence, instead of problem (\ref{eq:Q1}), 
one can address the problem of finding a partition $\p$ of
$G$ such that
\begin{equation}
|{\rm Cut}({\cal P})|-{\rm ExCut}({\cal P},{\cal G})
\rightarrow \min. \label{eq:version2}
\end{equation}

The last expression shows that we can solve (\ref{eq:Q1}) as a
problem of finding a MWC in a complete graph $G'$ with a vertex
set $V(G)$ and weight ${\rm weight}(i,j)$ on any edge $(i,j)\in
E(G')$ defined by
\begin{equation}
{\rm weight}(i,j) = \left\{
\begin{array}{l l l}
&  1 - p_{ij}, & \mbox{\rm  ~~~if }  (i,j)\in E(G) \\
  &  -p_{ij}, & \mbox{\rm  ~~~if }  (i,j)\not\in E(G),
\label{eq:weights}
\end{array}\right.
\end{equation}
where $p_{ij}$ is the probability that there is an edge
between vertices $i$ and $j$ in a random graph from the
class $\cal G$. Then, problem (\ref{eq:Q1}) is equivalent
to the problem of finding a partition $\cal P'$ of $G'$
such that
\begin{equation}
|{\rm Cut}({\cal P}')| \rightarrow \min.
\label{eq:version3}
\end{equation}

We summarize these observations in the following theorem.

\begin{theorem}
The problem of finding a partition of a given graph
$G=(V,E)$ that minimizes the modularity can be reduced in
$O(|V|+|E|)$ time to the problem of finding a minimum
weight cut in a complete graph $G'=(V,E')$ with edge
weights given by (\ref{eq:weights}).
\end{theorem}

For the reduction time bound in Theorem 1 we assume that
the edges of $E'\setminus E$ are defined implicitly. There
are several choices for $\cal G$ that have been favored by
various researchers. The random graph model $G(n,p)$ of
Erd\"{o}s-Renyi \cite{ErdosRenyi} defines $n$ vertices and
puts an edge between each pair with probability $p$.
Clearly, the expected number of edges of $G(n,p)$ is ${n
\choose 2} p$. Hence, for a graph with
expected number of edges $m$ %
\begin{equation}
p_{ij}=p=\frac{m}{{n \choose 2}}\;\cdot \label{eq:Gnp}
\end{equation}

One disadvantage of the $G(n,p)$ model 
is that it fails to capture important features of the
real-world networks, in particular, the degree
distribution. As has been recently observed
\cite{BarabasiAlbert}, many important types of networks
like technological networks (the Internet, the WWW), social
networks (collaboration networks, online social networks),
biological networks (protein interactions) have degree
distributions that follow a \emph{power law}, e.g., the
fraction of the vertices that have degree $k>0$ is roughly
proportional to $\alpha k^{-\lambda}$ for some constants
$\alpha$ and $\lambda>0$. Such networks are called
\emph{scale-free}. In comparison, the degrees of a random
graph from the $G(n,p)$ model follow a Poisson
distribution, i.e., the probability that a given vertex
has degree $k$ is ${n \choose k}p^k(1-p)^{n-k}$
and the expected degree of each vertex is $pn$. Hence, the
Erd\"{o}s-Renyi model may not be suitable as a choice for
$\cal G$ when used for determining the community structure
of graphs of the above type.

One model that takes into account the degrees of the
vertices is studied by Chung and Lu in \cite{ChungLu}. In
that model, the probability that there is an edge between
a vertex $i$ and a vertex $j$ is
\begin{equation}
p_{ij}=\frac{d_id_j}{\sum_{k=1}^nd_k}\,,\label{eq:CL}
\end{equation}
where $d_1,\cdots,d_n$ are positive reals corresponding to
the degrees of the vertices such that $\max_{1\le i\le
n}d_i^2<\sum_{i=1}^nd_i$. (The last condition guarantees
that such a graph exists if all numbers $d_i$ are
integers.) We will refer to that model as the Chung-Lu (CL)
model. Clearly, in the CL model, the expected degree of
vertex $i$ is $d_i$, compared with $pn$ (i.e., independent
on $i$) in the $G(n,p)$ model.

In the next section we will describe an efficient method
for finding a MWC of a graph $G'$ with weights
on the edges satisfying (\ref{eq:weights}) and $p_{ij}$
defined by (\ref{eq:Gnp}) or (\ref{eq:CL}).

\subsection{Finding a MWC using multilevel graph partitioning}
\label{sec:multilevel}

Above we established an important relationship between the graph
clustering and the MWC problems, i.e., that the problem of
finding a partition of a given graph that maximizes the
modularity can be reduced to the problem of finding a minimum
weight cut. Most existing work on the MWC problem considers
the case where all weights are non-negative. The MWC problem in
the case of non-negative weights is known to be polynomially
solvable, e.g., by using algorithms for computing maximum flows
\cite{AMO93}. In contrast, the MWC problem in case of
real-value weights is NP-hard and there is very little known for the
general version of the problem. Here we show that available
heuristics for another related problem, graph partitioning, can
be adapted to solve this version of the MWC problem.

\subsubsection{Overview of the multilevel  partitioning method.}
\begin{figure}[ht]\label{fig:ML}
\centering
\includegraphics[height=3in]{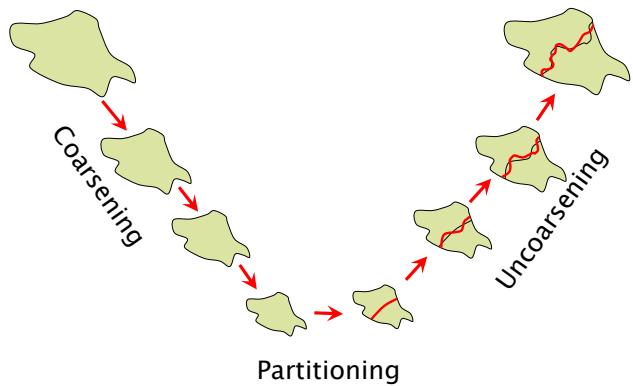}
\caption{The stages of multilevel partitioning.}
\end{figure}

Formally, the \emph{graph partitioning} (GP) problem is, given a
graph $G=(V,E)$, find a partition $(V_1,V_2)$ of $V$ such
that $||V_1|-|V_2||\leq 1$ (i.e., the partition is \emph{balanced})
and ${\rm Cut}(V_1,V_2)$ is
minimized. (Some versions of the problem consider
partitions of arbitrary cardinalities.) Note that, in comparison
with the minimum cut problem, there is the additional requirement
for a balanced partition. Because of its
important applications, e.g., in high performance
computing and VLSI design, GP is a well-researched
problem for which very efficient methods have been developed.
One such approach is the multilevel GP, which is
both fast and accurate for a wide
class of graphs that appear in practical applications.
Inspired by the multigrid method from computational
mathematics, it has been used in the works of Barnard and
Simon \cite{BarnardSimon}, Hendrickson and Leland
\cite{HendricksonLeland}, Karypis and Kumar
\cite{Metis1,Metis2}, and others. The method for bisecting
a graph consists of the following three phases(Figure~\ref{fig:ML}):

\smallskip\noindent
\emph{Coarsening phase}. 
The original graph $G$ is coarsened by partitioning it
into connected subgraphs and replacing each of the
subgraphs by a single vertex and replacing the set of the
edges between any pair of shrunk subgraphs by a single
edge. Moreover, a weight of each new vertex
(respectively edge) is assigned equal to the sum of the weights of
the vertices (respectively edges) that it represents. (Weights
on the original vertices of $G$ will be defined depending on whether the
$G(n,p)$ or the CL model has been used, as detailed below.) The
resulting graph is coarsened repeatedly by the same
procedure until one gets a graph of a sufficiently small
size. Let $G_0=G,G_1,\dots,G_l$ be the resulting graph
sequence.

\smallskip\noindent
\emph{Partitioning phase}. 
The graph $G_l$ is partitioned into two parts using any
available partitioning method (e.g., spectral partitioning
or the Kernighan-Lin (KL) algorithm \cite{KL}).

\smallskip\noindent
\emph{Uncoarsening and refinement phase.} 
The partition of $G_l$ is projected on $G_{l-1}$. Since the
weight of each vertex of $G_l$ is a sum of the weights of
the corresponding vertices of $G_{l-1}$, then the partition
of $G_{l-1}$ will be balanced if the partition of $G_l$ is
and the cut of both partitions will have the same weight.
However, since $G_{l-1}$ has more vertices than $G_l$, it
has more degrees of freedom and, therefore, it is possible
to refine the partition of $G_{l-1}$ in order to reduce its
cut size. For this end, the projection of the partition of
$G_l$ is followed by a refinement phase, which is usually
based on the KL algorithm. In the same way, the resulted
partition of $G_{l-1}$ is converted into a partition of
$G_{l-2}$ and refined, and so on until a partition of $G_0$
is found.

\medskip
\noindent \textbf{Kernighan-Lin refinement.} Since the
refinement step is the most involved part of the algorithm,
and which ultimately determines its accuracy and efficiency,
we will describe it in more detail. It has been shown \cite{Metis2} that the
KL algorithm can be a good choice for performing the
refinement.

The KL algorithms involves several iterations, each
consisting of moving a vertex from one set of the
partition to the other. Let ${\cal P}=\{P_1,P_2\}$ be the current
partition. For each vertex $u$ of the graph
a \emph{gain} for $u$ is defined as %
\begin{equation}
gain(u)=\sum_{v\in N(u)\setminus
P(u)}weight(u,v)-\sum_{v\in N(u)\cap P(u)}weight(u,v),
\label{eq:gain}
\end{equation}
where $N(u)$ is the set of all neighbors of $u$ and $P(u)$
is that set of $\cal P$ that contains $u$. $gain(u)$
measures how the weight of the cut will be affected if $u$
is moved from $P(u)$ to the other set of $\cal P$.
The KL algorithm then selects a vertex $w$ from the
smaller set of the partition with a maximum gain, moves it
to the other set, and updates the gains of the vertices
adjacent to $w$. Moreover, $w$ is marked so that it will
not be moved again during that refinement step. The
process is continued until either all vertices have been
moved, or the 50 most recent moves have not led to a
better partition. At the end of the refinement step, the
last $s\le 50$ moves that have not improved the partition
are reversed.


\medskip
\noindent \textbf{Implementation.} The implementation of
our algorithm for clustering is based on the version of
multilevel partitioning implemented by Karypis and Kumar
\cite{Metis1,Metis2}, which has been made freely available
as a software package under the name METIS.
Note that graph
partitioning, minimum cut, and clustering are related, but
with important differences problems, as illustrated in
Table~\ref{tab:comp}. We already showed how the clustering
problem can be reduced to a minimum cut problem and here we
will show how the resulting minimum cut problem can be
solved by a graph partitioning algorithm based on METIS.
Because of the differences between graph partitioning and MWC,
we have to make some evident changes.
For instance, since graph partitioning requires balanced partitions,
we have to drop the requirement for
balance of the partition. We have also to determine the
cardinality of the partition that minimizes the cut size.
But the main implementation difficulty is related to the
size of $G'$. Although the original graph, $G$, is
typically sparse, i.e., has $n$ vertices and $O(n)$ edges,
the transformed one, $G'$, is always dense, as it has
${n\choose 2}=\Omega(n^2)$ edges. The main challenge will
be to construct an algorithm whose complexity is close to
linear on the size of the original graph, rather than on
the size of the transformed one. We have shown that it is
possible to simulate an execution of the multilevel
algorithm on $G'$ by explicitly maintaining information
only about the edges from the original graph $G$ and
implicitly taking into account the remaining edges by
modifying the formulae for computing weights and gains. %
For instance, if ${\cal P}=\{P_1,P_2\}$ is a partition of
$V(G)$ and we have computed the value of the cut ${\rm cut}(P_1,P_2)$
of $G$ corresponding to $\cal P$ and maintain the values of $n_1=|P_1|$
and $n_2=|P_2|$, then the cut in $G'$ corresponding to $\cal P$ is
$${\rm cut}(P_1,P_2)-n_1n_2p$$ in the case of the $G(n,p)$ model
and hence can be computed in $O(1)$ time. A similar formula
holds for the case of the CL model.

\begin{table}[t]
  \centering
\begin{tabular}{|c||c|c|c|}
\hline
Problem & Clustering & Minimum Cut & Graph Partitioning \\
\hline \hline
Objective & Minimize modularity & Minimize cut size & Minimize cut size\\
\hline
Balance of partition & Sizes may differ & Sizes may differ & Equal sizes\\
\hline
Cardinality of partition & To be computed & To be computed & An input parameter\\
\hline
\end{tabular}
\smallskip
  \caption{Comparison between the clustering, minimum cut, and partitioning problems.}
  \label{tab:comp}
  \vspace*{-0.7cm}
\end{table}

\subsubsection{Clustering into an optimal number of clusters.}

The algorithm described above is a \emph{bisection} algorithm, i.e., it finds a partition (and
hence clustering) of the input graph into two parts. Our
algorithm for an arbitrary number of clusters
%
%
%
uses the following recursive procedure. We run the
bisection algorithm described above and let ${\cal P}$ be
the resulting partition. If ${\cal P}$ consists of only
one set (i.e., the original graph $G$ does not have a good
cluster partition), we are done. Else, we run recursively
the bisection algorithm on the two subgraphs $G_1$ and
$G_2$ of $G$ induced by the vertices of the two sets of
${\cal P}$. It is important to keep, during that recursive
call, the weights of the edges computed during the first
iteration instead of recomputing them based on $G_1$ and
$G_2$. The reason is that the random graph model based on
$G$ will be different than those based on $G_1$ and $G_2$
since formulae (\ref{eq:Gnp}) and (\ref{eq:CL}) will
produce different values for $p_{ij}$. It can be proven
that, if the bisection algorithm finds a minimum bisection
cut, the recursive algorithm described above finds a
minimum cut (of any number of parts) and hence finds a
clustering maximizing the modularity.

\vspace*{-0.2cm}
\subsubsection{Time analysis.} By
using
the analysis of Fiduccia and Mattheyses of the KL
algorithm from \cite{FM}, it follows that clustering any
network of $n$ vertices and $m$ edges into two communities
by our algorithm takes $O(n\log n+m)$ time, where $n$ and
$m$ are the numbers of the nodes and links of the network,
respectively. Finding a clustering in optimal number of
$k$ parts takes $O((n\log n+m)d)$ time, where $d$ is the
depth of the dendrogram describing the clustering
hierarchy. Although the worst-case value of $d$ can be
$\Omega(k)$, typically $d=O(\log k)$ \cite{CNM}.

\section{Experiments}

We performed a number of experiments on randomly generated
graphs in order to measure the accuracy of our algorithm
and its efficiency as well as to compare it with previous
algorithms. We chose Newman-Girvan algorithm
\cite{NewmanGirvan} and Clauset-Newman-Moore algorithm
\cite{CNM} since they are considered one of the best
existing algorithms and because of the code availability.

\subsection{Comparison with Newman-Girvan algorithm}
Following the experimental setting of \cite{NewmanGirvan},
we generated random graphs with 128 vertices and 4
communities of size 32 each. The expected degree of any
vertex is 16, but the \emph{outdegree} (the expected
number of neighbors of a vertex that belong to a different
community) is set to $i$ in the $i$-th experiment ($i\le
16$). Hence, higher values of $i$ correspond to graphs
with weaker cluster structures. The experiment is intended
to measure the sensitivity of the algorithm to the quality
of clustering.

\begin{table}
  \centering
\begin{tabular}{|c|c|c|c|}
\hline
 Outdegree  & Degree& Newman-Girvan & Ours \\ \hline
 1  & 16 & 1.00 & 1.00 \\ \hline
 2  & 16 & 1.00 & 1.00 \\ \hline
 3  & 16 & 0.98 & 0.99 \\ \hline
 4  & 16 & 0.97 & 0.99 \\ \hline
 5  & 16 & 0.95 & 0.99 \\ \hline
 6  & 16 & 0.85 & 0.97 \\ \hline
 7  & 16 & 0.60 & 0.91 \\ \hline
 8  & 16 & 0.30 & 0.70 \\ \hline
\end{tabular}
\medskip
  \caption{Comparing the quality of the clustering of our algorithm and \cite{NewmanGirvan}.}
  \label{table:exp1quality}
\end{table}

Table~\ref{table:exp1quality} compares the quality of the
clusterings produced by Newman-Girvan's algorithm and ours.
A clustering produced by any of the algorithms is
considered "correct" if it matches the original partition
of communities from the graph generation phase. (Note that,
due to the probabilistic nature of the graphs, the
clustering that maximizes the modularity might be different
from the original partition, especially if the modularity
is low.)

Our algorithm classifies correctly more than 99\% of the
edges for outdegrees $0, 1, 2, 3, 4, 5$ and in all cases it
is better than Newman-Girvan's (more than twice better for
the case $i=8$).

\subsection{Comparison with Clauset-Newman-Moore algorithm}
Table \ref{tab:compCNM} compares the performance of our
algorithm with Clauset, Newman, and Moore's algorithm
\cite{CNM}. That algorithm has the same quality of the
clustering as \cite{Newman2004}, but is claimed to be much
faster.  The test graphs in all experiments are random
graphs with different number of clusters, sizes,
densities, and modularities. Each experiment has been run
100 times on different random graphs.

In experiments 1--15 the random
graphs were generated in the following way: a graph with no
edges is created whose vertices are divided into subsets that
correspond to the clusters; then edges are created with
probability $p_{in}$ between vertices in the same subset and
with
probability $p_{out}$ between vertices from different subsets.
Experiments 1--8
compare how the performance of the algorithms depends on
the number of clusters, which vary from 2 to 9. The
results indicate that our algorithm produces always
clusterings with better quality, and the difference
increases when the number of the clusters grows. In
experiments 9--12 the test graphs have the same number of
vertices, number of cluster, and modularity, but different
densities. Those experiments show that our algorithm is
more sensitive when the density decreases, and in all the
cases our algorithm performs better. In experiments 13--15, we compare
the algorithms when the modularity (the quality of the
original clustering) is very low. We determined that with
modularity less than approximately 0.15 the algorithm from
\cite{CNM} is better, and if the modularity is greater
than 0.15 our algorithm is better. In all the above experiments, the
running time of our algorithm is considerably smaller,
whereby our algorithm is between 7 and 30 times faster
than the algorithm from \cite{CNM}.

Finally, in experiments
16--21 the random graphs were created such that their expected degree
sequences satisfy a power law distribution. The exponent of the
density function varies from -1.0 in experiment 16 to -2.0 in
experiment 21 in increments of -0.2. The results of the
experiments imply that in the case of power-law degree
distributions (scale-free graphs)
the quality of our algorithm consistently beats the one of the
algorithm from \cite{CNM}, while our time is in average 54
times smaller than theirs.

\begin{table}[t]
  \centering
\begin{tabular}{|c||c|c|c|c|c|c|c|c|c|}
\hline
Exp. & \# vert. &    \# edges & \# clust. & $Q_{\rm {orig}}$ & $Q_{\rm {CNM}}>$ & $Q_{\rm {ours}}>$ & $Q=$& $T_{\rm {CNM}}$ & $T_{\rm {ours}}$ \\
\hline \hline
1 & 200 & 8930 & 2 & .388 & 0 & 8 & 92 & .61 & .03  \\
\hline
2 & 300 & 14891 & 3 & .466 & 0 & 22 & 78 & 1.01 & .05 \\
\hline
3 & 400 & 21853 & 4 & .474 & 0 & 42 & 58 & 1.24 & .11 \\
\hline
4 & 500 & 29801 & 5 & .463 & 0 & 57 & 43 & 1.71 & .23 \\
\hline
5 & 600 & 38776 & 6 & .446 & 0 & 70 & 30 & 2.25 & .15 \\
\hline
6 & 700 & 48706 & 7 & .426 & 1 & 87 & 12 & 2.90 & .22 \\
\hline
7 & 800 & 59666 & 8 & .406 & 2 & 96 & 2 & 3.71 & .33 \\
\hline
8 & 900 & 71546 & 9 & .387 & 1 & 99 & 0 & 4.44 & .35 \\
\hline \hline
9 & 200 & 9932 & 2 & .298 & 0 & 8 & 92 & .68 & .04 \\
\hline
10 & 200 & 4967 & 2 & .299 & 0 & 27 & 73 & .54 & .03 \\
\hline
11 & 200 & 2458 & 2 & .298 & 0 & 50 & 50 & .61 & .02 \\
\hline
12 & 200 & 1238 & 2 & .295 & 6 & 92 & 2 & .46 & .00 \\
\hline \hline
13 & 400 & 41856 & 4 & .176 & 32 & 63 & 5 & 1.61 & .18 \\
\hline
14 & 400 & 43607 & 4 & .154 & 39 & 60 & 1 & 1.66 & .10 \\
\hline
15 & 400 & 47797 & 4 & .122 & 89 & 11 & 0 & 1.84 & .07 \\
\hline \hline
16 & 400 & 8537 & 4 & .244 & 0 & 100 & 0 & 1.35 & .02 \\
\hline
17 & 400 & 4879 & 4 & .273 & 0 & 100 & 0 & 1.33 & .01 \\
\hline
18 & 400 & 2653 & 4 & .308 & 0 & 100 & 0 & 1.33 & .03 \\
\hline
19 & 400 & 1449 & 4 & .370 & 0 & 100 & 0 & 1.36 & .04 \\
\hline
20 & 400 & 888 & 4 & .375 & 0 & 100 & 0 & 1.35 & .02 \\
\hline
21 & 400 & 629 & 4 & .394 & 0 & 100 & 0 & 1.34 & .03 \\
\hline
\end{tabular}
\medskip
  \caption{Comparison between the performances of our algorithm and \cite{CNM}.
$Q_{\rm {orig}}$ is the modularity of the partition used
during graph generation, "$Q_{\rm {CNM}}>$", "$Q_{\rm
{ours}}>$", and "$Q=$" are the percentages of the cases
where the algorithm \cite{CNM} produced a better
modularity, our algorithm produced a better modularity, or
both algorithms produced equal modularities, respectively.
$T_{\rm {CNM}}$ and $T_{\rm {ours}}$ are the times of the
algorithm from \cite{CNM} and ours, respectively.}
  \label{tab:compCNM}
\end{table}

\subsection{Testing on real-world data graphs}
\begin{figure}[ht]
\centering
\includegraphics[height=2.2in]{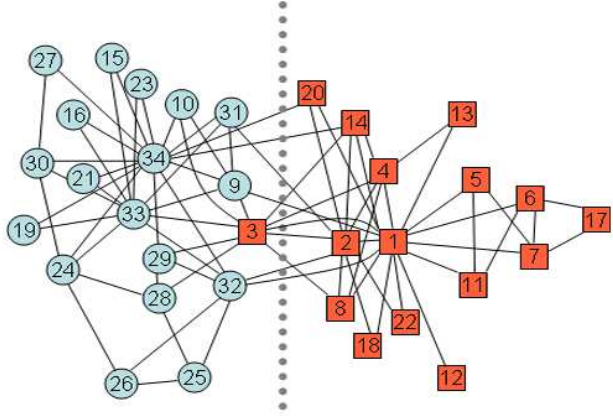}
\caption{Zachary's karate club network. Members of the communities
resulting after the split are denoted by circles and squares, respectively.
The communities found by our algorithm are separated by the vertical line.}
\label{fig:zachary}
\end{figure}

We tested our algorithms on a number of real-world graphs such
as the \emph{nd.edu} domain data \cite{AJB}, the United States college football
data \cite{GN2002}, and the Zachary's karate club
network \cite{Zachary}. In all cases our algorithm produced
clustering consistent with our previous knowledge of the
communities. For example, we describe in more detail here the
Zachary club network. This example is a standard benchmark for
community detection algorithms, describing the interactions between the
members of a karate club, which consequently split into two
because of  between the members, thereby revealing the hidden
communities of the original network. As shown on
Figure~\ref{fig:zachary}, our algorithm classified correctly
the members of the two subgroups, except for node 10. That node
has the same number of links (five) to both communities, hence
adding it to the smaller community results in a greater
modularity (e.g., our partitioning has a better modularity than
the "real" one.)

\subsection{Measuring the scalability}
\begin{table}[ht]
  \centering
\begin{tabular}{|c|c|c|c|c|c|}
\hline ${\rm p_{in}}$ & ${\rm p_{out}}$  &  Vertices  &
Edges &  Total size & Time (sec.) \\ \hline 0.10  &  0.01
&  5,000 &  406,125 & 411,125 & 1.77 \\ \hline 0.14  &
0.01  &  6,000 &  764,126 & 770,126 & 3.09 \\ \hline 0.18
&  0.01  &  7,000 &  1,283,398 &  1,300,398 &  3.22 \\
\hline 0.20  &  0.011 &  8,000 &  1,863,710 &  1,871,710 &
6.66 \\ \hline 0.20  &  0.013 &  9,000 &  2,418,730 &
2,427,730 &  5.68 \\ \hline 0.21  &  0.014 &  10,000 &
3,153,106 &  3,163,106 &  7.27 \\ \hline 0.22  &  0.015 &
15,000 & 6,295,801 &  6,310,801 &  15.18 \\ \hline
\end{tabular}
\medskip \caption{Measuring the scalability of our
algorithm. $p_{in}$ (respectively $p_{out}$) is the
expected fraction of the number of in-cluster (respectively
between cluster) edges to the number of all pairs of
vertices from the same set ( respectively different sets)
of the partition used for graph generation.}
  \label{table:exp1speed}
\end{table}

We also tested the speed of our algorithms by running them
on a 2 GHz desktop computer on graphs of different sizes.
The results are illustrated on
Table~\ref{table:exp1speed} and clearly show the
extraordinary speed and scalability of our algorithms.

\section{Conclusion}
This paper proposes a new approach for graph clustering by
reducing the clustering problem to a minimum cut problem
and then solving the latter problem by applying methods
for graph partitioning. Our proof-of-concept
implementation, based on the METIS partitioning package,
demonstrated the practicality of the approach. The changes
we made to METIS were minimal and various improvements and
refinements that take into account the specifics of the
clustering problem, use alternative minimum cut or graph
partitioning algorithms, or apply heuristics and parameter
adjustments in order to improve the accuracy are possible
and will be topics of further research.

\bigskip\noindent \textbf{Acknowledgement.} The author is
indebted to Melih Onus for helping with the programming
and most of the experiments and for many helpful
discussions.  We also would like to thank the developers of
METIS for making their source code publicly available.

\end{document}